\documentclass[10pt]{article}
\usepackage{graphicx} 
\usepackage{cite}
\usepackage{amsmath}
\usepackage{amssymb}
\usepackage{multirow}
\usepackage{adjustbox}
\title{Cosmological singularities in non-canonical models of dark energy}
\author{Oem Trivedi$^a$ \footnote{oem.t@ahduni.edu.in}  \and Simran Kaur Saggu $^{b}$ \footnote{sunkaur241@gmail.com} \and Pankaj Joshi $^{a}$ \footnote{psjcosmos@gmail.com}  }
\date{%
	$^a$International Centre for Space and Cosmology, School of Arts and Sciences, Ahmedabad University, Ahmedabad-380009, Gujarat, India\\%
	$^b$International Center for Cosmology, Charusat University, P.D.patel Institute of Applied sciences, Anand 388421 (Gujarat),India\\%
	\today
}
\begin{document}

\maketitle

\begin{abstract}

    The pursuit of unraveling the true essence of dark energy has become an immensely captivating endeavor in modern cosmology. Alongside the conventional cosmological constant approach, a diverse range of ideas has been proposed, encompassing scalar field-based models and various modified gravity approaches. A particularly intriguing notion involves exploring scalar field dark energy models within quantum gravitationally motivated cosmologies, with non-canonical theories standing out as a prominent candidate in this context. Hence, in this work, we investigate three widely recognized non-canonical scalar field dark energy models: phantom, quintom, and DBI dark energy models. By employing the Goriely-Hyde procedure, we demonstrate the presence of singularities in both finite and infinite time within these frameworks. and that these singularities can manifest regardless of the system's initial conditions. Moreover, we further establish how cosmological singularities of types I-IV can arise in all of these models. The work goes to show that non-canonical regimes for dark energy can allow for most of the prominent cosmological singularities for a variety of models. 
\end{abstract}

\section{Introduction}

	Observations of late-time acceleration of the Universe took the cosmological community by surprise \cite{SupernovaSearchTeam:1998fmf}. Since then, significant efforts have been made to explain this expansion, including the standard approaches like the Cosmological constant \cite{SupernovaSearchTeam:1998fmf,Weinberg:1988cp,Lombriser:2019jia,Copeland:2006wr,Padmanabhan:2002ji}, as well as more exotic scenarios like Modified gravity theories \cite{Capozziello:2011et,Nojiri:2010wj,Nojiri:2017ncd}, and recent proposals for direct detection of dark energy \cite{Zhang:2021ygh}. One fascinating approach to understanding dark energy is Quintessence, where a scalar field drives the late-time cosmic acceleration of the universe \cite{Zlatev:1998tr, Tsujikawa:2013fta,Faraoni:2000wk,Gasperini:2001pc,Capozziello:2003tk,Capozziello:2002rd,Carroll:1998zi,Caldwell:2005tm,Han:2018yrk,Astashenok:2012kb,Shahalam:2015sja,Oikonomou:2015qha}. Quintessence is particularly interesting as it represents the simplest scalar field dark energy scenario that does not suffer from issues like ghosts or Laplacian instabilities. In quintessence models, a slowly varying scalar field with a potential $V(\phi)$ leads to the acceleration of the universe, similar to the mechanism of slow-roll inflation. However, in this case, contributions from non-relativistic matter, such as baryons and dark Matter, cannot be ignored.
\\
\\
It is worth noting that simple models of Quintessence have been shown to be in conflict with the current H0 tension \cite{Colgain:2019joh,Banerjee:2020xcn,DiValentino:2021izs}, suggesting that simple Quintessence models may perform worse than $\Lambda$-CDM models in light of the current H0 data \cite{riess2021comprehensive}. This leads one to consider other more exotic possibilities for scalar field dark energy models and one such possibility is to consider models with noncanonical Lagrangians. Non-canonical models of scalar fields present some non-trivial characteristics, as they are subject to theoretical issues that do not arise in canonical cases. These issues include the presence of ghosts and nonphysical solutions. Nevertheless, the specific form of these contributions can be motivated by high-energy phenomenology, establishing a connection between cosmological observations and high-energy physics. As a result, non-canonical models have garnered significant attention in the literature on dark energy and modified gravity \cite{copeland2006dynamics,clifton2012modified,Papanikolaou:2022did,Orjuela-Quintana:2021zoe}. Three very interesting models of dark energy with noncanonical scalar fields are the phantom, quintom and Dirac-Born-Infeld (DBI) models. Phantom models of dark energy refer to models with the equation of state parameter $w < -1$ which is known as the phantom regime \cite{caldwell2002phantom,Caldwell:2005tm}. A dark energy model able to produce an EoS with values below -1 has generated considerable interest from a phenomenological point of view and it is also not completely ruled out by observations \cite{aghanim2020planck}. One can write the action of a phantom field minimally coupled to gravity as \begin{equation}
    S = \int d^4 x \sqrt{-g}  \left[ \frac{1}{2}  (\nabla \phi)^2  - V(\phi) \right]
\end{equation} 
where $\partial \phi^2 = g^{\mu \nu} \partial_{\mu} \phi \partial_{\nu} \phi $ $V(\phi)$ is a self interacting potential. The Lagrangian under consideration exhibits phantom behavior, where its equation of state (EOS) can go below -1. In contrast, quintessence models describe dark energy with an EOS greater than -1. Consequently, it is not possible to cross the phantom barrier using a single canonical or phantom scalar field alone. However, an intriguing model that allows for such a crossing was proposed by Feng et al. \cite{feng2005dark}. This particular dark energy scenario features an EOS larger than -1 in the past and less than -1 in the present, consistent with current observations. While this can be accomplished with more general non-canonical scalar fields, the simplest model entails a quintom Lagrangian consisting of two scalar fields. The action for such a quintom fields minimally coupled to gravity can be written as \begin{equation}
    S = \int d^4 x \sqrt{-g} \left[ - \frac{1}{2} (\nabla \phi)^2 + \frac{1}{2} \partial \sigma^2 - V(\phi,\sigma) \right]
\end{equation}
where $\phi$ is a canonical scalar field, $\sigma$ is a non canonical phantom field and $V(\phi,\sigma)$ is a general potential for both fields. Finally, another interesting class of noncanonical scalar field models is the DBI model. The simplest form of the DBI Lagrangian for a scalar field theory can be given by \begin{equation}
    \mathcal{L}_{t} = V(\phi) \sqrt{1 + \partial \phi^2}
\end{equation}    
This form of the DBI Lagrangian was used to predict tachyons in low energy EFT descriptions of string theory \cite{sen2002field,green1988witten} and since then Tachyonic scalar fields have also been studied in the context of cosmology \cite{padmanabhan2002accelerated,gibbons2002cosmological}. The general action for a minimally coupled DBI field, however, is an extension of (3) given by \cite{silverstein2004scalar} \begin{equation}
    S = \int d^4 x \sqrt{-g} \left[ \frac{1}{f(\phi)} (\sqrt{1 + 2 f(\phi) X  }) -1 ) - V(\phi) \right] 
\end{equation}
where $X = - \frac{1}{2} g^{\mu \nu} \partial_{\mu} \phi \partial_{\nu} \phi $ with $f(\phi) $ and $V(\phi)$ are arbitrary functions of $\phi$. These models have also been extensively studied in the context of cosmology have even been extended further \cite{bahamonde2018dynamical,burrage2014screening}. 
\\
\\
In recent times, a substantial body of literature has emerged that focuses on investigating the different types of cosmological singularities that may arise in the current and far future of the Universe \cite{Nojiri:2004ip,Nojiri:2005sr,Nojiri:2005sx,Bamba:2008ut,Bamba:2010wfw,Nojiri:2008fk,odintsov2022did,Trivedi:2023zlf,Odintsov:2023weg,nojiri2010future,Bahamonde:2016wmz,Bamba:2011sm,Odintsov:2015zza,Bamba:2012ka,Frampton:2011rh,Brevik:2011mm,Nojiri:2003vn,Briscese:2006xu,Abdalla:2004sw,Nojiri:2003ft,Elizalde:2004mq,Nojiri:2004pf,deHaro:2023lbq}. But often it is very difficult to classify and study the cosmological singularities which may occur in extremely non-conventional cosmologies which are motivated by quantum gravitational/ phenomenological considerations (for example, see the classification of singularities in asymptotically safe cosmology \cite{Trivedi:2022svr} ) and often it may not even be possible to do so in an orthodox fashion. Hence it becomes essential to look for non-conventional ways to find out cosmological singularities in exotic cosmologies and in this regard, a particular dynamical systems method can be of huge help. From a dynamical standpoint, one of the most intriguing aspects of studying various dynamical systems lies in understanding their singularity structure, which becomes particularly relevant when these systems describe physically significant phenomena. While numerous approaches have been proposed to explore the singularity structure of autonomous dynamical systems, one particularly interesting method is the Goriely-Hyde procedure \cite{goriely2000necessary}. As cosmology presents a multitude of captivating dynamical systems \cite{bahamonde2018dynamical}, the investigation of singularity structure in such systems has gained considerable attention, with the Goriely-Hyde method proving particularly useful for cosmological explorations \cite{barrow2004more,cotsakis2007dominant,cotsakis2007asymptotics,antoniadis2010brane,antoniadis2013brane,antoniadis2014enveloping}. This method has previously been applied to study finite and non-finite time singularities in certain classes of quintessence models as well \cite{odintsov2018dynamical,odintsov2019finite,Trivedi:2022ipa}. However, a comprehensive study of cosmological singularities in non-canonical scalar field models of dark energy using this approach is still lacking, and thus, we aim to address this gap in our work. In Section II, we provide a concise overview of the Goriely-Hyde method, after which we apply the complete Goriely-Hyde procedure to phantom, quintom, and DBI models of dark energy, as discussed previously. We demonstrate how singularities in these non-canonical models can exhibit diverse characteristics and occur at both finite and infinite times in Section III. Subsequently, in Section IV, we consider two well-motivated ansatz for the Hubble parameter and classify which types of cosmological singularities (Types I-IV) can arise within these regimes. Finally, we conclude our work in Section V.
\\
\\
	 \section{Goriely-Hyde Procedure}  
	 The Goriely-Hyde method \cite{goriely2000necessary} provides an elegant approach to determining the presence of finite-time singularities in dynamical systems. The procedure can be outlined as follows:

\begin{itemize}
	\item We begin by considering a dynamical system described by $n$ differential equations of the form:
	\begin{equation}
	\dot{x}_{i} = f_{i}(x),
	\end{equation}
	where $i = 1, 2, ..., n$, and the overdot represents differentiation with respect to time $t$, which in the case of quintessence models can be better represented by the number of e-foldings $N$. We identify the parts of the equation $f_{i}$ that become significant as the system approaches the singularity. These significant parts are referred to as "dominant parts" \cite{goriely2000necessary}. Each dominant part constitutes a mathematically consistent truncation of the system, denoted as $\hat{f}_{i}$. The system can then be written as:
	\begin{equation}
	\dot{x}_{i} = \hat{f}_{i}(x).
	\end{equation}
	
	\item Without loss of generality, the variables $x_{i}$ near the singularity can be expressed as:
	\begin{equation}
	x_{i} = a_{i} \tau^{p_{i}},
	\end{equation}
	where $\tau = t - t_{c}$, and $t_{c}$ is an integration constant. Substituting equation (4) into equation (3) and equating the exponents, we can determine the values of $p_{i}$ for different $i$, which form the vector $\mathbf{p} = (p_{1}, p_{2}, ..., p_{n})$. Similarly, we calculate the values of $a_{i}$ to form the vector $\vec{a} = (a_{1}, a_{2}, ..., a_{n})$. It is important to note that if $\vec{a}$ contains only real entries, it corresponds to finite-time singularities. Conversely, if $\vec{a}$ contains at least one complex entry, it may lead to non-finite-time singularities. Each set $(a_{i}, p_{i})$ is known as a dominant balance of the system.
	
	\item Next, we calculate the Kovalevskaya matrix given by:
	\begin{equation}
	R = \begin{pmatrix}
	\frac{\partial f_{1}}{\partial x_{1}} & \frac{\partial f_{1}}{\partial x_{2}} & . & . & \frac{\partial f_{1}}{\partial x_{n}}\\
	\frac{\partial f_{2}}{\partial x_{1}} & \frac{\partial f_{2}}{\partial x_{2}} & . & . & \frac{\partial f_{2}}{\partial x_{n}}\\
	. & . & . & . & . \\
	. & . & . & . & . \\
	\frac{\partial f_{n}}{\partial x_{1}} & \frac{\partial f_{n}}{\partial x_{2}} & . & . & \frac{\partial f_{n}}{\partial x_{n}}\\
	\end{pmatrix} -  \begin{pmatrix}
	p_{1} & 0 & . & . & 0 \\
	0 & p_{2} & . &

 . & 0 \\
	. & . & . & . & . \\
	. & . & . & . & . \\
	0 & 0 & . & . & p_{n} \\
	\end{pmatrix}.
	\end{equation}
	
	After obtaining the Kovalevskaya matrix, we evaluate it for different dominant balances and determine the eigenvalues. If the eigenvalues are of the form $(-1, r_{2}, r_{3}, ..., r_{n})$, with $r_{2}, r_{3}, ... > 0$, then the singularity is considered general and will occur regardless of the initial conditions of the system. Conversely, if any of the eigenvalues $r_{2}, r_{3}, ...$ are negative, the singularity is considered local and will only occur for certain sets of initial conditions.
\item Without loss of generality, the variables $x_{i}$ near the singularity can be expressed as:
\begin{equation}
x_{i} = a_{i} \tau^{p_{i}},
\end{equation}
where $\tau = t - t_{c}$, and $t_{c}$ is an integration constant. Substituting equation (4) into equation (3) and equating the exponents, we can determine the values of $p_{i}$ for different $i$, which form the vector $\mathbf{p} = (p_{1}, p_{2}, ..., p_{n})$. Similarly, we calculate the values of $a_{i}$ to form the vector $\vec{a} = (a_{1}, a_{2}, ..., a_{n})$. It is important to note that if $\vec{a}$ contains only real entries, it corresponds to finite-time singularities. Conversely, if $\vec{a}$ contains at least one complex entry, it may lead to non-finite-time singularities. Each set $(a_{i}, p_{i})$ is known as a dominant balance of the system.

\item Next, we calculate the Kovalevskaya matrix given by:
\begin{equation}
R = \begin{pmatrix}
\frac{\partial f_{1}}{\partial x_{1}} & \frac{\partial f_{1}}{\partial x_{2}} & . & . & \frac{\partial f_{1}}{\partial x_{n}}\\
\frac{\partial f_{2}}{\partial x_{1}} & \frac{\partial f_{2}}{\partial x_{2}} & . & . & \frac{\partial f_{2}}{\partial x_{n}}\\
. & . & . & . & . \\
. & . & . & . & . \\
\frac{\partial f_{n}}{\partial x_{1}} & \frac{\partial f_{n}}{\partial x_{2}} & . & . & \frac{\partial f_{n}}{\partial x_{n}}\\
\end{pmatrix} -  \begin{pmatrix}
p_{1} & 0 & . & . & 0 \\
0 & p_{2} & . & . & 0 \\
. & . & . & . & . \\
. & . & . & . & . \\
0 & 0 & . & . & p_{n} \\
\end{pmatrix}.
\end{equation}

After obtaining the Kovalevskaya matrix, we evaluate it for different dominant balances and determine the eigenvalues. If the eigenvalues are of the form $(-1, r_{2}, r_{3}, ..., r_{n})$, with $r_{2}, r_{3}, ... > 0$, then the singularity is considered general and will occur regardless of the initial conditions of the system. Conversely, if any of the eigenvalues $r_{2}, r_{3}, ...$ are negative, the singularity is considered local and will only occur for certain sets of initial conditions.

\end{itemize} 
	\section{Singularity Analysis}
	 \subsection{Phantom Dark Energy}
Assuming a flat FLRW metric form for the cosmology, we can write and using the Lagrangian (1), we can write the cosmological equations in the presence of a matter fluid given \footnote{Throughout this paper we are working in a usual flat FLRW background i.e. the universe that we consider is the usual isotropic and homogeneous one. Future explorations in this direction can indeed make one to consider, for example, how these singularities would occur in an anisotropic universe with let's say the Bianchi universe ( Bianchi type I for simplicity ) or even a Kantowski-Sachs cosmology in the case that one wants to take into account more exotic phenomenological features. But that is for now beyond the scope of this work as we would like to explore the singularities in non-canonical theories in rather simple settings first.} by $p_{m} = w_{m}\rho_{m}$ ($0 \leq w_{m} \leq \frac{1}{3}$)  as \cite{bahamonde2018dynamical} \begin{equation}
    3 H^2 = \kappa^2 \left( \rho_{m} - \frac{\dot{\phi}^2}{2} + V(\phi) \right)
\end{equation}
\begin{equation}
    2 \dot{H} + 3 H^2 = \kappa^2 \left( -w \rho + \frac{1}{2} \dot{\phi}^2 + V(\phi) \right)
\end{equation}
While the Klein-Gordon equation has its usual form \begin{equation}
    \ddot{\phi} + 3 H \dot{\phi} + V^{\prime} (\phi) = 0  
\end{equation}
where the overdots denote differentiation with respect to the coordinate time and the prime denotes differentiation with respect to $\phi$. The EOS of the field is given by \begin{equation}
    w_{\phi} = \frac{\frac{1}{2} \dot{\phi}^2 + V(\phi) }{\frac{1}{2} \dot{\phi}^2 - V(\phi) } 
\end{equation}
with the energy density of the scalar field given by \begin{equation}
    \rho_{\phi} = - \frac{1}{2} \dot{\phi}^2 + V(\phi) 
\end{equation}
One sees immediately that assuming a positive scalar field potential when the kinetic energy equals the potential energy the EoS (14) diverges. This can be taken as a first warning that the theoretical pathologies mentioned above can yield non-physical behavior \footnote{A comment is in order about the issues of a phantom cosmological field. Changing the sign of the kinetic energy unavoidably results in an unbounded total energy of the scalar field \cite{carroll2003can,cline2004phantom}. From a quantum perspective, this leads to the emergence of ghost modes that violate unitarity, while from a classical perspective, solutions of the equations of motion become unstable under small perturbations and violate the dominant energy condition \cite{carroll2003can,cline2004phantom}. Due to these significant concerns, some view the phantom scalar field as a simplistic phenomenological model, considering it as an emergent phenomenon that should not be trusted at the deepest fundamental level. Nevertheless, these concerns do not discourage the exploration of singularities within this regime.}. In order to kick start our analysis, we define the variables \footnote{Note that it is not necessary that the variables will be defined in this same way for all cosmological paradigms, as we shall see later in the paper too. In fact, one can use different variables for the same paradigm too if required or wished for. See, for example, \cite{halliwell1987scalar,copeland1998exponential,faraoni2013scalar} for extended discussions on the same} to be   \begin{equation}
    x = \frac{\kappa \dot{\phi} }{\sqrt{6} H} \qquad y = \frac{\kappa \sqrt{V}}{\sqrt{3 }H} \qquad \lambda = - \frac{V^{\prime} }{V}
\end{equation} 
Using this one can write the cosmological equations (11-13) as \begin{equation}
    x^{\prime} = \frac{1}{2} \left( 3 (w-1) x^3 - 3x \bigg[ y^2 + 1 + w(y^2 - 1 )  \bigg] - \sqrt{6} \lambda y^2 \right)  
\end{equation}
\begin{equation}
    y^{\prime} = - \frac{1}{2} y \Bigg[ - 3 ( w- 1) x^2 + 3 ( w + 1 ) (y^2 - 1 ) + \sqrt{6} \lambda x \Bigg]
\end{equation}
\begin{equation}
    \lambda^{\prime} = - \sqrt{6} (\Gamma - 1 ) x \lambda^2 
\end{equation}
where $$ \Gamma = \frac{V V^{\prime \prime} }{{V^{\prime}}^{2}} $$ and the primes on the variables denote differentiation with respect to $ \eta = \log a $. It is important to note that (17-19) do not form an autonomous system unless $\Gamma$ can be written as a function of $\lambda$, in which case it forms a 3D autonomous system. While it is interesting to study this system for a variety of potentials, we shall focus here on the case of exponential potential only. In this case, $\lambda$ takes the form of a constant and (17-18) form a 2D autonomous system and hence fit to be dealt with the Goriely-Hyde procedure. This particular form of potential is also arguably the most studied form for this model \cite{hao2003attractor,urena2005scalar} and thus is also indeed very popular. 
Now that we have decided on our system (17-18), we can start our analysis. The first truncation that we consider is given by \begin{equation}
         \hat{f} = \begin{pmatrix}
         -\left(\frac{3}{2} w x y^2\right)  \\
         \frac{1}{2} y \left(3 (1-w) x^2\right) 
         \end{pmatrix}
         \end{equation}
Using the ansatz for the Goriely-Hyde method described in section II, we get the exponent values as $\mathbf{p} = (-1/2,-1/2) $ and using this we get \begin{equation}
         \begin{aligned}
         a_{1} = \left(\frac{i}{\sqrt{3 (1-w)}} , \frac{1}{\sqrt{3 w}} \right) \\[10pt]
         a_{2} = \left(\frac{i}{\sqrt{3 (1-w)}} , \frac{-1}{\sqrt{3 w}} \right) \\[10pt]
         a_{3} = \left(\frac{-i}{\sqrt{3 (1-w)}} , \frac{1}{\sqrt{3 w}} \right) \\[10pt]
         a_{4} = \left(\frac{-i}{\sqrt{3 (1-w)}} , \frac{-1}{\sqrt{3 w}} \right)
         \end{aligned}
         \end{equation}
         as all $\mathbf{\hat{a}}$ have complex entries \footnote{At this point we would like to highlight that complex entries in $\mathbf{\hat{a}}$  and those observed in (15) are completely consistent with the fact that the system we have considered in (11-13) consists of x,y and z which are real and positive. As mentioned in section II, complex entries for various $\mathbf{a}$ suggest that the singularities will be non-finite time in nature and hence these quantities taking up complex values is consistent with the analysis as shown in \cite{goriely2000necessary}. Similar case has been for various cosmological systems (for example, see \cite{odintsov2018dynamical,odintsov2019finite})} , only non-finite time singularities will be possible with regards to this truncation. The Kovalevskaya matrix now takes the form \begin{equation}
             R = \left(
\begin{array}{cc}
 \frac{1}{2}-\frac{3 w y^2}{2} & -3 w x y \\
 3 (1-w) x y & \frac{3}{2} (1-w) x^2+\frac{1}{2} \\
\end{array}
\right)
        \end{equation}
After plugging in the dominant balances (21) into the Kovalevskaya matrix, we find its eigenvalues to be \begin{equation}
    r = (-1,-1)
\end{equation} 
Hence the singularities in this case will only be local singularities which will only form for a limited set of initial conditions. Furthermore, as the dominant balance (21) had complex entries we can say that this truncation also tells that the singularities could happen at non finite times while being very dependent on the initial conditions for the system variables (17-18). Note that we have not set any constraint on the value of $\lambda$ and so it does not matter what value $\lambda$ takes in this case.  
\\
\\
The second truncation that we consider is given by \begin{equation}
         \hat{f} = \begin{pmatrix}
         -\frac{1}{2} \left(\sqrt{6} \lambda  \text{y1}^2\right)  \\
         -\frac{1}{2} \left(\sqrt{6} \lambda  \text{x1} \text{y1}\right) 
         \end{pmatrix}
         \end{equation}
Using the ansatz for the Goriely-Hyde method described in section II, we get the exponent values as $\mathbf{p} = (-1,-1) $ and using this we get \begin{equation}
         \begin{aligned}
         a_{1} = \left( \frac{\sqrt{2}}{ \sqrt{3}\lambda } , \frac{\sqrt{2}}{ \sqrt{3}\lambda } \right) \\[10pt]
         a_{2} = \left(\frac{\sqrt{2}}{ \sqrt{3}\lambda } , - \frac{\sqrt{2}}{ \sqrt{3}\lambda }  \right)
         \end{aligned}
         \end{equation}
Here we immediately note that both $a_{1} $ and $a_{2} $ have only real entries, which means that the singularities occurring in this scenario can take place in finite time, something which we couldn't see in the previous truncation that we considered. Moving forward, we can write the Kovalevskaya matrix in this case to be \begin{equation}
    R = \left(
\begin{array}{cc}
 1 & -\sqrt{6} \lambda  \text{y1} \\
 -\sqrt{\frac{3}{2}} \lambda  \text{y1} & 1-\sqrt{\frac{3}{2}} \lambda  \text{x1} \\
\end{array}
\right)
\end{equation}
The eigenvalues of this after plugging in the dominant balance (25) is found to be \begin{equation}
    r = (-1,1)
\end{equation}
This tells that this truncation allows for the system to have general singularities to, singularities which will take place irrespective of the initial conditions for the system variables. This coupled with the conclusion we found by the dominant balances having real entries entails us to the fact that the phantom scenario described by (17-18) will surely have cosmological singularities occurring in both finite and nonfinite times regardless of the initial conditions. To classify exactly which singularities these would be physically would be done in section IV. In passing it is also worth mentioning that phantom dark energy models seem to be providing quite optimistic views towards possible solutions of prevalent cosmic tensions today, in particular the Hubble tension \cite{Alestas:2020mvb,Pan:2019gop,DiValentino:2020naf,Vagnozzi:2019ezj,DiValentino:2022eot,Escudero:2022rbq,Chudaykin:2022rnl,Adil:2023exv,DiValentino:2016hlg,Vagnozzi:2023nrq}. In general it is being seen that phantom-like behaviour of the effective equation of state of dark energy can more naturally accommodate higher values of $H0$, preferred by recent local measurements, while satisfying the CMB constraints as well. Hence one can see that there is a renewed sense of realism attached with phantom cosmological scenarios given tensions like this and it may very well be the key to alleviating the $H0$ issues once and for all.
\\
\\
\subsection{Quintom Dark Energy}
We now turn our attention toward the interacting quintom model described by (2). In this case, we consider the potential to be of the form \begin{equation}
    V(\phi,\sigma) = V_{0} \exp^{-\kappa \mathbf{(\lambda_{\phi} \phi + \lambda_{\sigma} \sigma )}}
\end{equation}
where $\lambda_{\phi}$ and $\lambda_{\sigma} $ are some constants. We consider this potential form because, again, the exponential potential has been the most studied type in this regard and this has displayed a variety of cosmologically interesting behaviors \footnote{Another reason to consider the coupled potential is that it is mathematically simpler to analyze dynamically than the uncoupled one,
because it only requires one expansion normalized variable for the potential energy rather than two \cite{bahamonde2018dynamical} as we shall see later in this section.} ( for example, \cite{lazkoz2006quintom} obtained tracking, phantom, and quintessence solutions for this potential ). The cosmological equations for this potential given the Lagrangian (2) are \begin{equation}
    3 H^2 = \kappa^2 \Bigg[ \rho_{m} + \frac{1}{2} \dot{\phi}^2 - \frac{1}{2} \dot{\sigma}^2 + V(\phi,\sigma) \Bigg]
\end{equation} 
\begin{equation}
    2 \dot{H} + 3 H^2 = -\kappa^2 \Bigg[ w_{m} \rho_{m} + \frac{1}{2} \dot{\phi}^2 - \frac{1}{2} \dot{\sigma}^2 - V(\phi,\sigma) \Bigg]
\end{equation}
While the Klein-Gordon equations for both the fields are given by \begin{equation}
    \ddot{\phi} + 3 H \dot{\phi} + \frac{\partial V}{\partial \phi} = 0 
\end{equation}
\begin{equation}
    \ddot{\sigma} + 3 H \dot{\sigma} - \frac{\partial V}{\partial \sigma} = 0 
\end{equation}
The expansion normalized variables for this system can be written as \begin{equation}
    x_{\phi} = \frac{\kappa \dot{\phi}}{\sqrt{6} H} \qquad x_{\sigma} = \frac{\kappa \dot{\sigma}}{\sqrt{6} H} \qquad y = \frac{\kappa \sqrt{V}}{\sqrt{3} H}
\end{equation}
Using these variable, one can write the equations (29-32) in the form of 3D autonomous system as \begin{equation}
    x^{\prime}_{\phi} = \frac{1}{2} \Bigg[ 3 x_{\phi} \bigg[ (w-1) x_{\sigma}^2 - (w+1) y^2 w - 1  \bigg]- 3 (w-1) x_{\phi}^3 + \sqrt{6} \lambda_{\phi} y^2  \Bigg]
\end{equation}
\begin{equation}
    x^{\prime}_{\sigma} = \frac{1}{2} \Bigg[ 3 (w -1) x_{\sigma}^3 - 3 x_{\sigma} \bigg[ (w-1 ) x_{\phi}^2  + (w+1) y^2  - w + 1  \bigg] - \sqrt{6} y^2 \lambda_{\sigma}  \Bigg]
\end{equation}
\begin{equation}
    y^{\prime} = - \frac{1}{2} y \Bigg[ 3 (w-1)(x_{\phi}^2 - x_{\sigma}^2 ) + 3 (w +1 ) ( y^2 -1 ) + \sqrt{6} \lambda_{\sigma} x_{\sigma} + \sqrt{6} \lambda_{\phi} x_{\phi} \Bigg]
\end{equation}
The eqns. (34-36) form a 3D autonomous system and so is fit to be studied with the Goriely-Hyde procedure. The first truncation that we consider is given by \begin{equation}
         \hat{f} = \begin{pmatrix}
         -\frac{1}{2} \left(3 (w-1) x_{\phi}^3\right)  \\
         -\frac{1}{2} \left(3 (w+1) x_{\sigma} y^2\right) \\ 
         \frac{3}{2} (w-1) x_{\sigma}^2 y
         \end{pmatrix}
         \end{equation}
Using the ansatz for the Goriely-Hyde method described in section II, we get the exponent values as $\mathbf{p} = (-1/2,-1/2,-1/2) $ and using this we get \begin{equation}
         \begin{aligned}
         a_{1} = \left( \frac{1}{\sqrt{3 (w-1)}} , \frac{i}{\sqrt{3 (w-1)}} , \frac{1}{\sqrt{3 (w+1)}} \right) \\[10pt]
         a_{2} = \left( - \frac{1}{\sqrt{3 (w-1)}} , \frac{i}{\sqrt{3 (w-1)}} , \frac{1}{\sqrt{3 (w+1)}}  \right) \\[10pt]
         a_{3} = \left(\frac{1}{\sqrt{3 (w-1)}} , - \frac{i}{\sqrt{3 (w-1)}} , \frac{1}{\sqrt{3 (w+1)}}  \right) \\[10pt]
         a_{4} = \left(\frac{1}{\sqrt{3 (w-1)}} ,  \frac{i}{\sqrt{3 (w-1)}} , - \frac{1}{\sqrt{3 (w+1)}}  \right) \\[10pt] 
         a_{5} = \left( \frac{1}{\sqrt{3 (w-1)}} ,- \frac{i}{\sqrt{3 (w-1)}} , -\frac{1}{\sqrt{3 (w+1)}} \right) \\[10pt]
         a_{6} = -\left( \frac{1}{\sqrt{3 (w-1)}} ,- \frac{i}{\sqrt{3 (w-1)}} , \frac{1}{\sqrt{3 (w+1)}} \right) \\[10pt]
         a_{7} = -\left( \frac{1}{\sqrt{3 (w-1)}} , \frac{i}{\sqrt{3 (w-1)}} , -\frac{1}{\sqrt{3 (w+1)}} \right) \\[10pt]
         a_{8} = \left(- \frac{1}{\sqrt{3 (w-1)}} ,- \frac{i}{\sqrt{3 (w-1)}} , -\frac{1}{\sqrt{3 (w+1)}} \right) 
         \end{aligned}
         \end{equation}
         as all $\mathbf{\hat{a}}$ have complex entries \footnote{At this point we would like to highlight that complex entries for the dominant balance in (38) and those observed in the previous section for phantom models in (21) are completely consistent with the fact that the system we have considered in for quintom and phantom models, (34-36) and (17-18) respectively, consists of the expansion normalized variables which are real and positive (x,y for phantom and $x_{\phi} $, $x_{\sigma} $ and y for quintom models) . As mentioned in section II, complex entries for various $\mathbf{\hat{a}}$ suggest that the singularities will be a non-finite time in nature and hence these quantities taking up complex values is consistent with the analysis as shown in \cite{goriely2000necessary}. A similar case has been for various cosmological systems (for example, see \cite{odintsov2018dynamical,odintsov2019finite})}, only non-finite time singularities will be possible with regards to this truncation. The Kovalevskaya matrix in this case takes the form \begin{equation}
             R = \left(
\begin{array}{ccc}
 \frac{1}{2}-\frac{9}{2} (w-1) x_{\phi}^2 & 0 & 0 \\
 0 & \frac{1}{2}-\frac{3}{2} (w+1) y^2 & -3 (w+1) x_{\sigma} y \\
 0 & 3 (w-1) x_{\sigma} y & \frac{3}{2} (w-1) x_{\sigma}^2+\frac{1}{2} \\
\end{array}
\right)
        \end{equation}
Using the dominant balance for this system in (38), we can find the eigenvalues of this Kovalevskaya matrix to be \begin{equation}
    r = \left( -1 , -1 , 1  \right) 
\end{equation}
We see here that one of the eigenvalues besides -1 is also negative and so as far as this truncation is concerned, one can only have local singularities for this cosmology. Coupling this with the form of the dominant balances (38), one sees that the Goriely-Hyde procedure tells us here that singularities can form in non-finite times for a particular set of initial conditions for the variables.
\\
\\
The second truncation that we consider is given by \begin{equation}
         \hat{f} = \begin{pmatrix}
         -\frac{1}{2} \left(3 (w+1) x_{\phi} y^2\right)  \\
         -\frac{1}{2} \left(3 (w-1) x_{\phi}^2 x_{\sigma} \right) \\ 
         -\frac{1}{2} \left(\sqrt{6} \lambda  x_{\sigma} y\right)
         \end{pmatrix}
         \end{equation}
Using the ansatz for the Goriely-Hyde method described in section II, we get the exponent values as $\mathbf{p} = (-1/2,-1,-1/2) $ and using this we get \begin{equation}
         \begin{aligned}
         a_{1} = \left( \frac{\sqrt{2}}{\sqrt{3 (w-1)}} , \frac{1}{\sqrt{6} \lambda } , \frac{1}{\sqrt{3 (w+1)}} \right) \\[10pt]
         a_{2} = \left( -\frac{\sqrt{2}}{\sqrt{3 (w-1)}} , \frac{1}{\sqrt{6} \lambda } , \frac{1}{\sqrt{3 (w+1)}}  \right)
         \end{aligned}
         \end{equation}
We note here that both $a_{1} $ and $a_{2} $ have only real entries, which means that the singularities occurring in this scenario can take place in finite time, something which we couldn't see in the previous truncation that we considered. Moving forward, we can write the Kovalevskaya matrix in this case to be  
   \begin{equation}
             R = \left(
\begin{array}{ccc}
 \frac{1}{2}-\frac{3}{2} (w+1) y^2 & 0 & -3 (w+1) x_{\phi} y \\
 -3 (w-1) x_{\phi} x_{\sigma} & 1-\frac{3}{2} (w-1) x_{\phi}^2 & 0 \\
 0 & -\sqrt{\frac{3}{2}} \lambda  y & \frac{1}{2}-\sqrt{\frac{3}{2}} \lambda  x_{\sigma} \\
\end{array}
\right)
        \end{equation}
Using the dominant balance for this system in (38), we can find the eigenvalues of this Kovalevskaya matrix to be \begin{equation}
    r = \left( -1 , 1/2 , 1/2  \right)  
\end{equation}
We see here that none of the eigenvalues besides -1 is negative and so this truncation tells us that general singularities not dependent on the initial conditions are also possible in this scenario. Coupled with the conclusions from the dominant balances, we see that the singularities in the case of the quintom models described by (34-36) can take place place in finite time and be independent of initial conditons. Note very interestingly that these singularities occur irrespective of whether or not the dark energy equation of state is in phantom ($w < -1 $) or quintessence regime ($w > -1$) and these results hold irrespective of the values of $\lambda_{\phi} $ and $\lambda_{\sigma} $, hence encapsulating a wide array of popular quintom models for cosmology \footnote{In principle one can make more interesting truncations for the quintom system like \begin{equation*}
         \hat{f} = \begin{pmatrix}
         \frac{3}{2} (w-1) x_{\phi} x_{\sigma}^2  \\
         \sqrt{\frac{3}{2}} \left(-\left( \lambda_{\sigma} y^2\right)\right) \\ 
         \sqrt{\frac{3}{2}} \lambda_{\phi} x_{\phi} (-y)
         \end{pmatrix}
         \end{equation*} 
But we choose to not pursue the analysis of more truncations as we have already demonstrated all the properties of singularities which can be known using the Goriely-Hyde method by using only two truncations. The reader is, however, welcome to analyze any more truncations if they wish to for their own interest. }. The physical classification of these singularities will take place in section 4 
\\
\\
\subsection{DBI Dark Energy}
We would like to consider the models corresponding to the action (4) now but before that, we would quickly like to discuss some subtleties of these form of DBI models. From the perspective of string theory, $ V (\phi) $ is a potential that arises from quantum interactions between a D3-brane associated with $\phi$ and other Dbranes. Although a quadratic potential was considered in the initial proposal of the generalized DBI model \cite{silverstein2004scalar}, discussions on the exact form of the potential are still ongoing. $f(\phi) $, on the other hand,  is the inverse of the D3-brane tension and contains geometrical information about the throat
in the compact internal space with some proposals for the form of $f(\phi)$ consider it to be a constant or even of the form $f(\phi) = \alpha \phi^{-4} $ ($ \alpha$ being a constant ), which is the case for an AdS throat. In the context of cosmology, it is normally taken to be reasonable to consider both $f(\phi) $ and $V(\phi) $ as non-negative \cite{bahamonde2018dynamical} and that is what we would also be considered in here. Another important parameter with regards to DBI models is the Lorentz factor given by \begin{equation}
    \gamma = \frac{1}{\sqrt{1 - f(\phi) \dot{\phi}^2 }}
\end{equation}
When the speed of the brane motion, $ |\dot{\phi} |  $, approaches the limit $f^{-1/2} $ the Lorentz factor can grow without bound and one starts to see the true effects of the DBI model. While in the limit when the brane motion can be neglected and is not comparable to the limits of $f^{-1/2} $, $\gamma \to 1$ and one sees the usual theory of canonical scalar fields unfold. The energy and pressure densities for the DBI field can be written using (4) as \begin{equation}
    \rho_{\phi} = 2 X \mathcal{L}_{,X} = \frac{\gamma -1 }{f(\phi)} + V(\phi)  
\end{equation}  
\begin{equation}
    p_{\phi} = \frac{\gamma -1 }{\gamma f(\phi) } - V(\phi)
\end{equation}
where $\mathcal{L}_{,X} $ refers to the Lagrangian's differentiation with respect to the kinetic term X. From this one can write the speed of sound for the DBI model as \begin{equation}
    c_{s}^2 = \frac{\partial p_{\phi} / \partial X}{\partial \rho_{\phi} / \partial X} = \frac{1}{\gamma^2} = 1 - f(\phi) \dot{\phi}^2  
\end{equation}
One thus sees that the speed of sound and the Lorentz factor are directly related to each other in DBI models. For a flat FLRW metric universe with scale factor $a(t)$, $ X = \dot{\phi}^2 / 2$ for a homogeneous field $\phi$ and one can then write the energy and pressure densities of such a field as \cite{copeland2010cosmological} \begin{equation}
    \rho_{\phi} = \frac{\gamma^2}{\gamma +1 } \dot{\phi}^2 + V
\end{equation} 
\begin{equation}
    p_{\phi} = \frac{\gamma^2}{\gamma +1 } \dot{\phi}^2 - V
\end{equation}
We can now write the Friedmann equation for this scenario as \begin{equation}
    \frac{3 H^2 }{\kappa^2} = \rho_{m} + \frac{\gamma^2}{\gamma +1 } \dot{\phi}^2 + V 
\end{equation}
The Klein-Gordon equation also gets significantly modified and takes the form \begin{equation}
    \ddot{\phi} + \frac{3 H}{\gamma^2} \dot{\phi} + \frac{1}{\gamma^3} \frac{dV}{d \phi} + \frac{1}{2f} \frac{df}{d \phi} \frac{(\gamma +2 ) (\gamma -1 ) }{(\gamma +1 ) \gamma } \dot{\phi}^2 = 0
\end{equation}
The continuity equation for the matter field still stays has the usual form \begin{equation}
    \dot{\rho}_{m} + 3 H \rho_{m} (1 + w_{m} ) = 0 
\end{equation}
Introducing now the following expansion normalized variables \cite{copeland2010cosmological,bahamonde2018dynamical}, \begin{equation}
    x = \frac{\gamma \kappa \dot{\phi}}{\sqrt{3 (\gamma +1 ) H}} \qquad y = \frac{\kappa \sqrt{V}}{\sqrt{3} H} \qquad \overline{\gamma} = \frac{1}{\gamma}
\end{equation}
Before proceeding further, we would like to specify that in our analysis we would like to consider DBI models with a constant $\gamma$ ( or equivalently, constant sound speed ) and this is motivated by two reasons. Firstly, we would like to base this analysis on rather simple models of DBI cosmology and would like to understand the extent to which cosmological singularities are allowed to occur in such models. When one considers models with $\gamma$ being dependent on $\phi$, one runs into quite a lot of complexities and as we just currently want to gauge the status quo of cosmological singularities in these models, for now we would like to first see out the arguably simpler DBI models( and in any case, models with constant $\gamma$ have not been short of attention in recent years by any means \cite{copeland2010cosmological}). Recently even structure formation has been studied comprehensively with constant $\gamma$ models \cite{fahimi2018structure}. A second reason is more subtle and concerns the very reason DBI models came to the fore very quickly ; non-Gaussianity. Although we are considering the very late universe in this work primarily, from the phenomenological viewpoint one of the main reasons why this model attracted strong attention is because of its sizable equilateral type of primordial non-Gaussianity. In particular, there are certain cases of constant $\gamma $ DBI models where one can hope to observe large yet detectable levels of non-Gaussianity in the early universe \cite{copeland2010cosmological,amani2018resurrecting,cai2009inflation,rasouli2019warm}. So while constant $\gamma$ models are interesting for us simply from the perspective that these are a simpler class of DBI models, they could have some deep implications for very interesting quantum gravitational results in particular epochs of the universe.
\\
\\
With that cleared, we can now write the cosmological equations (51-53) using the expansion normalized variables as \begin{equation}
 x^{\prime} = \frac{1}{2} \sqrt{3 \overline{\gamma } (\overline{\gamma} + 1 ) } \lambda y^2 + \frac{3}{2} x \Bigg[ (1+ \overline{\gamma } (x^2 -1 ) + (1+w ) (1- x^2 - y^2 ) \Bigg]   
\end{equation}
\begin{equation}
    y^{\prime} = - \frac{1}{2} \sqrt{3 \overline{\gamma} (1 + \overline{\gamma } } \lambda x y + \frac{3}{2} y \Bigg[ (1 + \overline{\gamma}) x^2 + (1+w ) (1- x^2 - y^2 ) \Bigg]
\end{equation}
where $\lambda = - \frac{1}{\kappa V} \frac{dV}{d \phi }$ is again considered to be a constant while we also considered another quantity, $\mu = - \frac{1}{k V} \frac{df}{d \phi}$ to be a constant as well. There is an important new feature that emerges in the DBI case and is not present in the case of the canonical scalar field. In the latter case only an exponential potential can truly lead to an autonomous system, but as was shown in \cite{copeland2010cosmological}, this fact does not apply to the DBI field and there can be non-exponential potentials which can still lead to constant $\lambda$ but here we assume an exponential form for the potential.
\\
\\
We can now start with the singularity analysis, with our first truncation being given by \begin{equation}
         \hat{f} = \begin{pmatrix}
         \frac{3}{2} (\overline{\gamma} +1) x^3  \\
         -\frac{1}{2} \left(3 (w+1) y^3\right) 
         \end{pmatrix}
         \end{equation}
Using the ansatz for the Goriely-Hyde method described in section II, we get the exponent values as $\mathbf{p} = (-1/2,-1/2) $ and using this we get \begin{equation}
         \begin{aligned}
         a_{1} = \left( \frac{i}{\sqrt{3 (\overline{\gamma} +1)}} , \frac{1}{\sqrt{3 (w+1)}} \right) \\[10pt]
        a_{2} = \left( -\frac{i}{\sqrt{3 (\overline{\gamma} +1)}} , \frac{1}{\sqrt{3 (w+1)}} \right) \\ 
         \end{aligned}
         \end{equation}
The Kovalevskaya matrix in this case is given by \begin{equation}
    R = \left(
\begin{array}{cc}
 \frac{9}{2} (\overline{\gamma} +1) x^2+\frac{1}{2} & 0 \\
 0 & \frac{1}{2}-\frac{9}{2} (w+1) y^2 \\
\end{array}
\right)
\end{equation}         
Using the dominant balance (58), we find that the eigenvalues of this matrix are given by \begin{equation}
    r = (-1,1)
\end{equation}
Hence the singularities in this case will only be local singularities which will only form for a limited set of initial conditions. Furthermore considering that the dominant balance for this truncation has complex entries, one can further predict from this truncation that the system admits singularities in non-finite time. 
\\
\\
The second truncation that we consider is given by \begin{equation}
         \hat{f} = \begin{pmatrix}
         \frac{1}{2} \sqrt{3 \overline{\gamma}  ( \overline{\gamma}   +1)} \lambda  y^2  \\
         \frac{1}{2} \sqrt{3 \overline{\gamma}  (\overline{\gamma} +1)} (-\lambda ) (x y) 
         \end{pmatrix}
         \end{equation}
Using the ansatz for the Goriely-Hyde method described in section II, we get the exponent values as $\mathbf{p} = (-1,-1) $ and using this we get \begin{equation}
         \begin{aligned}
         a_{1} = \left( \frac{2}{\sqrt{3 \overline{\gamma}  ( \overline{\gamma} +1)} \lambda } , \frac{2 i}{\sqrt{3 \overline{\gamma}  ( \overline{\gamma} +1)} \lambda } \right) \\[10pt]
         a_{2} = \left( \frac{2}{\sqrt{3 \overline{\gamma}  ( \overline{\gamma} +1)} \lambda } ,- \frac{2 i}{\sqrt{3 \overline{\gamma}  ( \overline{\gamma} +1)} \lambda } \right)
         \end{aligned}
         \end{equation}
Furthermore the Kovalevskaya matrix for this truncation can be written as \begin{equation}
    R = \left(
\begin{array}{cc}
 1 & \sqrt{3} \sqrt{ \overline{\gamma}  ( \overline{\gamma} +1)} \lambda  y \\
 -\frac{1}{2} \sqrt{3} \sqrt{\overline{\gamma}  ( \overline{\gamma} +1)} \lambda  y & 1-\frac{1}{2} \sqrt{3} \sqrt{ \overline{\gamma}  ( \overline{\gamma} +1)} \lambda  x \\
\end{array}
\right)
\end{equation}
Using the balance (62), we get the eigenvalues for this matrix to be \begin{equation}
r = (-1,2)    
\end{equation}
These eigenvalues directly tell us that the system allows for general singularities which are independent of the initial conditions. Coupled with the balance (62), one can conclude that according to the Goriely-Hyde method, this class of DBI models described by the system we have considered above will have singularities in any case irrespective of the initial conditions and that those singularities could occur at non-finite times. 
\\
\\
           \section{Physical Classification of the Singularities}
           Until now, we have discussed the singularity structure within this dark energy scenario from a dynamical perspective. However, it is insufficient to merely acknowledge the existence of singularities in this system from a physical standpoint. Thus, it becomes necessary to appropriately classify the potential types of singularities that could occur in this model. Various types of physical singularities for cosmology at a specific time $t = t_{s}$, where $t_{s}$ represents the occurrence of the singularities, can be classified as follows \cite{Nojiri:2005sx,Fernandez-Jambrina:2010ngm}:

\begin{itemize}
\item Type I ("Big Rip"): In this case, the scale factor $a$, effective energy density $\rho_{\text{eff}}$, and effective pressure density $p_{\text{eff}}$ diverge.
\item Type II ("Sudden/Quiescent singularity"): In this case, $p_{\text{eff}}$ diverges, as well as the derivatives of the scale factor beyond the second derivative.
\item Type III ("Big Freeze"): In this case, the derivative of the scale factor from the first derivative onwards diverges.
\item Type IV ("Generalized sudden singularities"): In this case, the derivative of the scale factor diverges from a derivative higher than the second.
\end{itemize}

Among these classifications, Type I singularities are considered strong singularities since they have the ability to distort finite objects, while singularities of Type II, Type III, and Type IV are regarded as weak singularities as they cannot be perceived as either the beginning or the end of the universe. Although there are other minor types of singularities, such as Type V singularities or "w" singularities, we will focus solely on Type I to Type IV singularities here. The most general form of the Hubble parameter for investigating singularities within the aforementioned classified types is expressed as \cite{odintsov2019finite}:

\begin{equation} \label{a11}
H(t) = f_{1}(t) + f_{2}(t)(t - t_{s})^{\alpha}
\end{equation}

Here, $f_{1}(t)$ and $f_{2}(t)$ are assumed to be nonzero regular functions at the time of the singularity, and similar conditions apply to their derivatives up to the second-order. Additionally, $\alpha$ is a real number. It is not mandatory for the Hubble parameter (34) to be a solution to the field equations; however, we will consider this case and explore the implications of this assumption on the singularity structure based on our dynamic analysis. First, we observe that none of the variables $x$, $y$, or $z$ as defined in (10) can ever become singular for any cosmic time value. The singularities that can occur considering the Hubble parameter as defined in (34) are as follows:

\begin{itemize}
\item For $\alpha < -1$, a big rip singularity occurs.
\item For $-1 < \alpha < 0$, a Type III singularity occurs.
\item For $0 < \alpha < 1$, a Type II singularity occurs.
\item For $\alpha > 1$, a Type IV singularity occurs.
\end{itemize}

Another ansatz useful for classifying singularities was introduced in \cite{odintsov2022did} whereby the scale factor was written as \begin{equation} \label{a12}
    a(t) = g(t) (t-t_{s})^{\alpha} + f(t)
\end{equation}
where g(t) and f(t) and all their higher order derivatives with respect to the cosmic time are smooth functions of the cosmic time. For this ansatz, according to the values of the exponent $\alpha$ one can have the following singularities \begin{itemize}
    \item For $\alpha < 0 $, a type I singularity occurs 
    \item For $0 < \alpha < 1$, a type III singularity develops
    \item For $a < \alpha < 2$, a type II singularity occurs 
    \item For $\alpha > 2$, a type IV singularity occurs 
\end{itemize}
Again, it is not mandatory that the scale factor in \eqref{a12} will necessarily be a solution to the field equations but we would like to consider this and \eqref{a11} in order to get a well-motivated feel for the type of cosmological singularities we can deal with in the various models we have discussed so far. 
\\
\\
To proceed further, we need to express the expansion normalized variables that we defined for the phantom, quintom, and DBI models in terms of the Hubble parameter alone. To do this, we realize that we need to express the potential and the derivative of the field parameter(s) in each case in terms of the Hubble parameter as these are the quantities on which the expansion normalized variables really depend in all models. For the phantom case given by the cosmological equations (11-12), one arrives at \begin{equation}
    \dot{\phi}^2 = \frac{2 \dot{H} }{k^2} + \rho_{m} (1-w_{m})
\end{equation} 
\begin{equation}
    V(\phi) = \frac{\dot{H} + 3 H^2 }{k^2} - \frac{\rho_{m}}{2} (1+w_{m}) 
\end{equation}
For the case of the Quintom model, one can write the interaction potential as \begin{equation}
    V(\phi,\sigma) = \frac{3 H^2 - \dot{H} }{\kappa^2} + \frac{\rho_{m}}{2} (w_{m} -1 ) 
\end{equation}
While writing separate expressions $\sigma$ and $\phi$ in the quintom case only in terms of the Hubble parameter and its derivatives can prove cumbersome, we can however write \begin{equation}
    \dot{\sigma}^2 - \dot{\phi}^2 = \rho_{m} (w_{m} + 1 ) + \frac{2 \dot{H}}{\kappa^2} 
\end{equation}
If the quantity $ \dot{\sigma}^2 - \dot{\phi}^2 $ is finite, we would be assured of both $\dot{\sigma}$ and $\dot{\phi} $ being finite as well and hence this expression would do the trick for us in this case. For the case of the DBI field, one can write \begin{equation}
    \dot{\phi}^2 = \frac{\gamma + 1}{\gamma -1 } \Bigg[ \rho_{m} (w_{m} -1 ) - \frac{2 \dot{H} }{\kappa^2} \Bigg]
\end{equation} 
\begin{equation}
    V(\phi) = \frac{3 H^2}{\kappa^2} - \rho_{m} + \frac{\gamma^2}{\gamma - 1 } \left( 
 \frac{2 \dot{H}}{\kappa^2} - \rho_{m} (w_{m} -1) \right) 
\end{equation}
Using these expressions, one can express the expansion normalized variables for each system as discussed before but as the expressions just become too long, we will not be putting them here and summarize our findings as follows.
\begin{table*}
\centering
\caption{Summary of Singularities for Different Ansatz}
\label{tab:singularities}
\begin{tabular}{|c|c|c|c|}
\hline
\multirow{2}{*}{Ansatz} & \multicolumn{3}{c|}{Singularities} \\ \cline{2-4} 
                        & Phantom & Quintom & DBI ($\gamma \neq 1$) \\ \hline
$H(t) = f_{1}(t) + f_{2}(t)(t - t_{s})^{\alpha}$ & Type I, II, III, IV & Type I, II, III, IV & Type I, II, III, IV \\ \hline
$a(t) = g(t) (t - t_{s})^{\alpha} + f(t)$ & None & Type IV & Type I, II, III, IV \\
(except in canonical limit) \\ \hline
\end{tabular}
\end{table*}
Table 1 explains the formation of various singularities in the regimes we have considered and we see that all types of cosmological singularities are allowed for the ansatz (65) in all the models we have considered, except for DBI model going to its canonical limit $\gamma \to 1$ in which case all of these singularities get removed ( but the model becomes canonical again ). The ansatz (66) provides a different picture as by this ansatz, one does not get any singularities in the phantom model, only gets type IV singularities in the quintom regime while the status quo of DBI models remains unchanged in that it still can have all the singularities until it does not go to its canonical limit. One is had to conclude from this that non-canonical regimes very easily seem to be allowing for a larger class of cosmological singularities than what has been seen through similar procedures for their canonical counterparts \cite{odintsov2018dynamical,odintsov2019finite, Trivedi:2022ipa}. Note that these results hold true independent of what is the equation of state of the background matter fluid $w_{m} $, meaning this is true for any configuration of matter background. Note also that in recent years some ideas have been discussed in the literature about ways in which the weaker cosmological singularities may be delayed or made milder, for example conformal anomalies have been seen to be helpful in this regard \cite{Nojiri:2004ip,Nojiri:1999mh}. But these anomaly driven effects also have some nuances attached to them, for example the quantum effective action in such cases (like the one considered here) is usually non-local and higher-derivative can lead to more serious issues near the time of the singularities and conformal anomaly effects are not always helpful for removing singularities \cite{Trivedi:2022svr}. Another way of removing the weaker singularities of the cosmological type that was considered was to introduce modified gravity effects, in particular f(R) effects \cite{Bamba:2008ut}. While such approaches to dealing with these singularities are certainly very appealing and could bear fruit, our work here is more focused on just finding out singularities and we see that a discussion on singularity removal methods in this scenario would be suitable for a whole new work altogether.
\\
\\
           \section{Concluding Remarks}
           In this paper, we have considered three very well-known noncanonical scalar field dark energy models in the form of phantom, quintom,
and DBI dark energy models. These three models have received immense attention from a phenomenological point of view in recent years and hence we found it interesting to probe cosmological singularities in these diverse non-canonical scenarios. For this endeavor, we used a method pioneered by the works of Odintsov in recent years, in which we applied the Goriely-Hyde procedure to the various dynamical systems by which the cosmological equations of these three models could be described. This allowed us to make predictions about whether singularities in these scenarios would be strongly dependent on initial physical conditions and whether they could happen in finite or non-finite times. After this, we employed two very well-motivated ansatz' for the Hubble parameter and the scale factor to reach 
the conclusion that one can have all of type I-type IV singularities in such cosmologies in certain cases. This work goes to show that non-canonical regimes for dark energy can allow for most of the prominent cosmological singularities for a variety of models. We find it very intriguing that in the models we have studied, the dark energy or the matter fields with negative pressures are not able to remove the space-time singularities. But this happens because sometimes the overall gravitational focusing of the energy density is more dominant as compared to the negative pressures. An interesting point to note is if all the singularities in the space time are weak then the space-time could be geodetically complete. In that case, then there are no genuine singularity in the space-time, even though some occasional blowups of some physical quantities may show up. On the other hand, when the singularity is strong, then there will be genuine incompleteness as for the geodesics, and in the case that we have studied here one sees that at least one strong singularity will occur (the big rip or type I singularity ). Overall, the ideas related to non-canonical models or similar phenomenologically inspired cosmological scenarios which have been introduced in cosmology for various reasons in literature are of very novel and rather uncertain nature.we know very little about their properties and hence examining the singularities caused by them could be one way of asking more and learning more about their properties. 
\\
\\
\section*{Acknowledgments}

The authors would like to thank Sergei Odintsov, and Alexander Timoshkin for very helpful discussions on the work and Maxim Khlopov for discussions on cosmological singularities in general. We would also like to thank the anonymous referees of the paper for their insightful comments on the work which have improved its depth multi-folds.  

\bibliography{citations}
\bibliographystyle{unsrt}

\end{document}